# Application and prospect of hydrogels in diabetic wound treatment

Jiayi Yang[1], Peiwu Qin[3*], Zhenglin Chen[2,3,4,5*]

[1]School of Public Health, Hebei Medical University, Shijiazhuang 050000, Hebei, China

[2]Key Laboratory of Imaging Diagnosis and Minimally Invasive Intervention Research, The Fifth Affiliated Hospital of Wenzhou Medical University, Lishui, Zhejiang Province, 323000, China.

[3]Tsinghua-Berkeley Shenzhen Institute, Tsinghua Shenzhen International Graduate School, Tsinghua University, Shenzhen 518000, Guangdong, China

[4]Institute of Biopharmaceutical and Health Engineering, Shenzhen International Graduate School, Tsinghua University, Shenzhen 518000, Guangdong, China

[5]Key Lab for Industrial Biocatalysis, Ministry of Education, Department of Chemical Engineering, Tsinghua University, Beijing 100084, China

* Corresponding author's emails: chenzlin1992@163.com

Abstract:

Diabetic foot patients face persistent, challenging wounds, prompting a quest for innovative therapies. Hydrogel, a promising modality, is increasingly recognized for its potential in diabetic wound care. This review examines hydrogel's emergence as a therapeutic option, and its benefits, and outlines research directions. Diabetic foot ulcers, notoriously hard to heal, demand effective solutions beyond traditional methods. Hydrogel, with its hydrophilic nature and biocompatibility, offers a novel approach. It fosters a moist wound environment, aids healing, and mitigates infection risks. Additionally, its versatility in delivering bioactive agents enhances wound recovery. This analysis provides insights into hydrogel's application, paving the way for improved outcomes and patient well-being in diabetic foot care.

1. Introduction

Diabetic foot patients often suffer from chronic wounds that are difficult to heal, posing a significant challenge in the medical field. The limitations of traditional treatment methods have led to an urgent need for innovative therapies. Hydrogel, as a novel treatment modality, has garnered attention in recent years and has shown potential in treating wounds in diabetic foot patients. This work aims to review the emergence of hydrogel as a new therapeutic modality in treating wounds in diabetic foot patients, explore its research purposes and significance, and outline the structure of this review.

Patients with diabetic foot often have chronic foot ulcers, which are not only difficult to heal but also prone to infection, pain, and even amputation. Traditional treatment methods, including local debridement, antibiotic use, and dressing coverage, have limited efficacy, leading to drug resistance and side effects, as well as long treatment cycles and high costs, imposing significant physical and mental burdens on patients.

Hydrogel is a type of gel material with hydrophilic properties, its excellent biocompatibility and controllable physicochemical properties make it a novel choice for treating wounds in diabetic foot patients. Hydrogel can provide a moist healing environment, promote wound healing, and possess good adsorption and moisturizing properties, effectively controlling wound exudate, reducing the risk of infection, and providing comfortable protection for pathological wounds. Moreover, hydrogel can also be used to carry biologically active substances, such as growth factors and antimicrobial agents, further promoting wound healing.

This review aims to systematically summarize the application and advantages of hydrogel in treating wounds in diabetic foot patients and analyze its research progress and future development trends in clinical practice. By understanding the characteristics, mechanisms, and clinical applications of hydrogel, more treatment options can be provided for healthcare workers, bringing better treatment outcomes and quality of life for diabetic foot patients.

2. Challenges in Treating Diabetic Foot Wounds

Diabetic foot patients often confront the challenge of non-healing foot ulcers, greatly impacting their life and health. This section will delve into the characteristics of diabetic foot wounds and the limitations of traditional treatment methods.

2.1. Characteristics of Diabetic Foot Wounds

Diabetic foot is characterized by foot lesions resulting from changes in nerves, blood vessels, and immune function due to diabetes, including foot ulcers, infections, gangrene, etc. The characteristics of wounds in diabetic foot patients are as follows: In terms of susceptibility to infection, damaged foot tissues due to neuropathy and impaired blood circulation increase the risk of infection, even minor trauma can lead to infection. Regarding healing, elevated blood glucose levels reduce tissue repair capacity, making foot ulcers difficult to heal, often occurring in high-pressure areas like toe clefts and soles, subjecting the healing process to continuous pressure and friction. In terms of neuropathy, patients often suffer from peripheral neuropathy, reducing pain sensation in foot wounds, delaying treatment, and potentially leading to foot deformities and abnormal walking, increasing the risk of foot ulcers. Regarding vascular changes, arterial sclerosis and microcirculatory disorders result in inadequate blood supply to foot tissues, affecting healing, and may lead to serious complications such as gangrene. Regarding multifactorial influences, the occurrence and development of diabetic foot are influenced by various factors, including hyperglycemia, infection, trauma, foot deformities, inappropriate footwear, etc., making treatment complex.

2.2. Limitations of Traditional Treatment Methods

Traditional treatment methods for diabetic foot wounds entail several limitations. Firstly, the treatment period is prolonged, requiring continuous procedures like regular debridement and dressing changes, extending treatment periods for weeks or even months. Secondly, due to the weakened immune function in diabetic foot patients, foot tissues are prone to infection, and traditional treatment methods often struggle to effectively control infections, leading to worsening wounds or complications. Moreover, traditional dressings often fail to alleviate patient pain effectively, especially in cases of larger or deeper wounds, pain often affects the patient's quality of life. Prolonged use of antibiotics may lead to bacterial resistance, reducing treatment efficacy, and some patients may experience allergic reactions to antibiotics, limiting their use. Additionally, the long treatment period and frequent medical procedures of traditional treatment methods increase the economic burden of treatment, particularly for economically disadvantaged patients, and treatment difficulties may arise. Finally, due to the unique nature of diabetic foot, wounds often recur, especially with improper treatment or inadequate wound management, recurrence is common, leading to repeated illness. In conclusion, traditional treatment methods have several limitations in treating diabetic foot wounds, necessitating the search for new treatment modalities to improve treatment outcomes and enhance patient quality of life. Hydrogel, as a novel treatment modality, holds immense potential to address this challenge, bringing new hope for the treatment of wounds in diabetic foot patients.

3. Basic Principles and Characteristics of Hydrogel

Hydrogel, as a novel biomaterial, has shown immense potential in the field of wound treatment. This section will introduce the basic principles and characteristics of hydrogel, including its definition and classification, mechanisms of action in wound treatment, as well as its advantages and features.

## 3.1. Definition and Classification of Hydrogel

Hydrogel is a gel material with high water absorption capacity, consisting mainly of water and polymers. Hydrogels exhibit strong water-absorbing properties, capable of adsorbing large amounts of water and forming a gel, with excellent biocompatibility and biodegradability. Depending on their composition, hydrogels can be classified into two types: natural hydrogels and synthetic hydrogels. Natural hydrogels mainly include gelatin, alginate, etc., while synthetic hydrogels are composed of synthetic polymers such as polyacrylic acid, polyvinyl alcohol, etc. Based on the different gel forms, hydrogels can be divided into solid hydrogels and sol-gel hydrogels. Solid hydrogels possess strong mechanical properties and stability, whereas sol-gel hydrogels form a gel-like substance in water, making them easy to apply and cover on the wound surface.

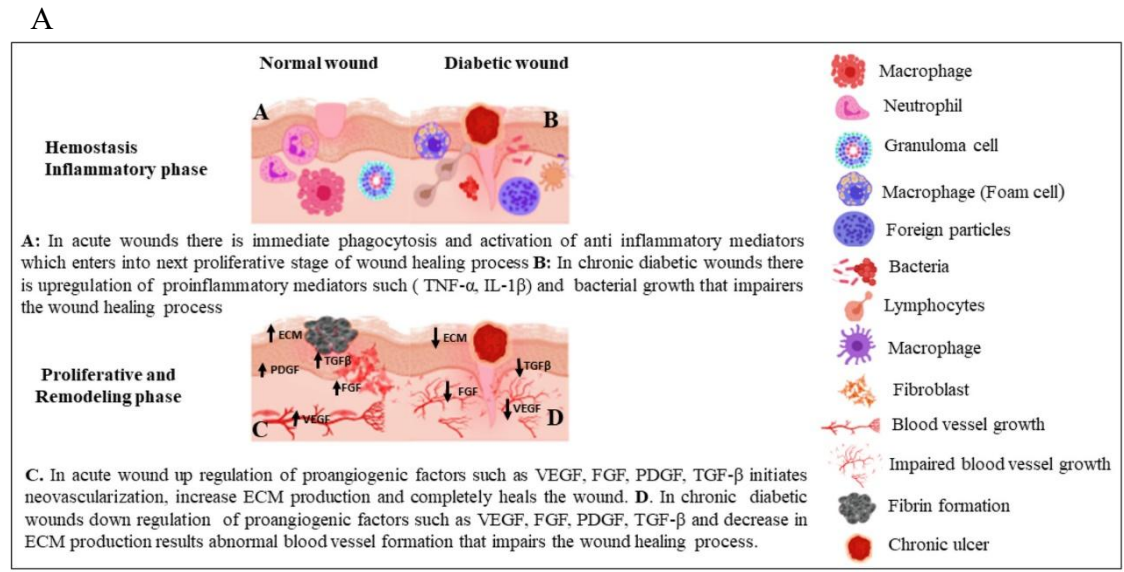

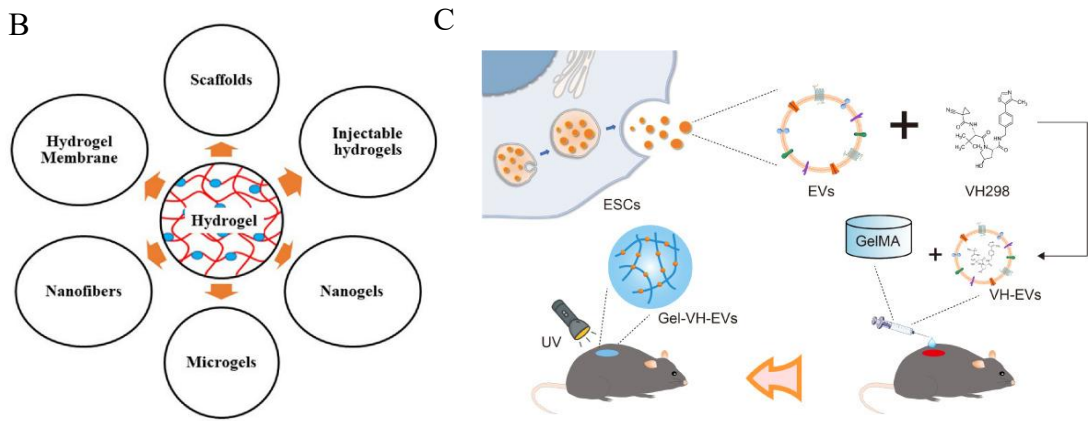

Fig.1 (A) Normal wound versus diabetic wound healing. Copyright@ [1] (B) Various types of hydrogels used in drug delivery and tissue engineering. Copyright@ [2] (C) The experimental procedure of the gelatin methacryloyl hydrogel for diabetic wound healing in vivo study. Copyright@ [3]

## 3.2. Mechanisms of Action of Hydrogel in Wound Treatment

Hydrogel, as a novel dressing for diabetic wound treatment, exhibits unique mechanisms in promoting wound healing. Its primary mechanisms include moisturizing, absorption and drainage, promotion of healing, protection, and inflammation regulation. Firstly, regarding moisturizing, hydrogel effectively maintains a moist environment on the wound surface, aiding in preventing skin dryness caused by high blood sugar and promoting the growth and regeneration of epithelial cells. Secondly, concerning absorption and drainage, hydrogel rapidly absorbs wound exudate and seals it within the gel, thereby preventing recontamination of the wound surface and reducing the risk of infection. Furthermore, in promoting healing, hydrogel contains various active ingredients such as growth factors and antimicrobials, stimulating the growth and repair of new tissue, and accelerating wound healing, while inhibiting infection and inflammatory responses. Additionally, hydrogel forms a soft protective film, effectively isolating external stimuli and contaminants, reducing bacterial and viral invasion, relieving pressure and friction around the wound, and protecting the wound from further damage. Lastly, in inflammation regulation, the active ingredients in hydrogel can modulate inflammatory responses, reducing the severity of inflammation, and aiding in wound healing and repair. for example, Mude et al. provided an overview of in situ gelling injectable hydrogels for diabetic wounds, emphasizing the mechanism of action of these hydrogels in treating diabetic wounds [1]. Gangwar et al. reviewed various novel drug delivery strategies for the treatment of bacterial skin and soft tissue infections, including hydrogel dressings [2]. Wang et al. reported on extracellular vesicles released from gelatin methacryloyl hydrogel for diabetic wound healing [3]. These studies collectively highlight the potential of hydrogels in wound treatment and provide insights into their mechanisms of action[4–34].

3.3. Advantages and characteristics of hydrogels

Hydrogel dressings have been developed as a modern wound treatment alternative for chronic diabetic wounds[4]. Research has also shown that hydrogels can be effective in promoting wound repair and regeneration, as evidenced by fibroblast migration to the wound site in a 3D type 2 diabetic human skin model[5]. Additionally, hydrogels have been loaded with therapeutic agents such as polydeoxyribonucleotide (PDRN) to accelerate diabetic wound healing in animal models [6]. The advances in hydrogel dressings in diabetic wounds have been systematically summarized in recent literature, providing theoretical support for devising hydrogel dressings and inspiration for diabetic wound treatment [38]. Furthermore, functional hydrogels have been explored as wound dressings to enhance wound healing, and pH/glucose dual-responsive hydrogel dressings have been developed for athletic diabetic foot wound healing [39]. Recent reviews have focused on the advancement of functional hydrogels for diabetic wound management, discussing the normal healing process and the role of macrophages in the process [40]. Overall, hydrogels have been shown to have many advantages and characteristics that make them promising for the treatment of diabetic wounds, including their ability to promote wound repair and regeneration, deliver therapeutic agents, and respond to specific physiological conditions[34,41–60].

4. Application of hydrogels in the treatment of diabetic wounds

Diabetic foot patients often face the challenge of chronic wounds that are difficult to heal, significantly impacting their quality of life and health. In recent years, hydrogel has garnered attention as a novel treatment modality for diabetic wound care. This review will delve into the role of hydrogel in promoting healing, reducing infection risk, alleviating pain, and improving quality of life in diabetic wound management, while also looking ahead to its prospects in this field.

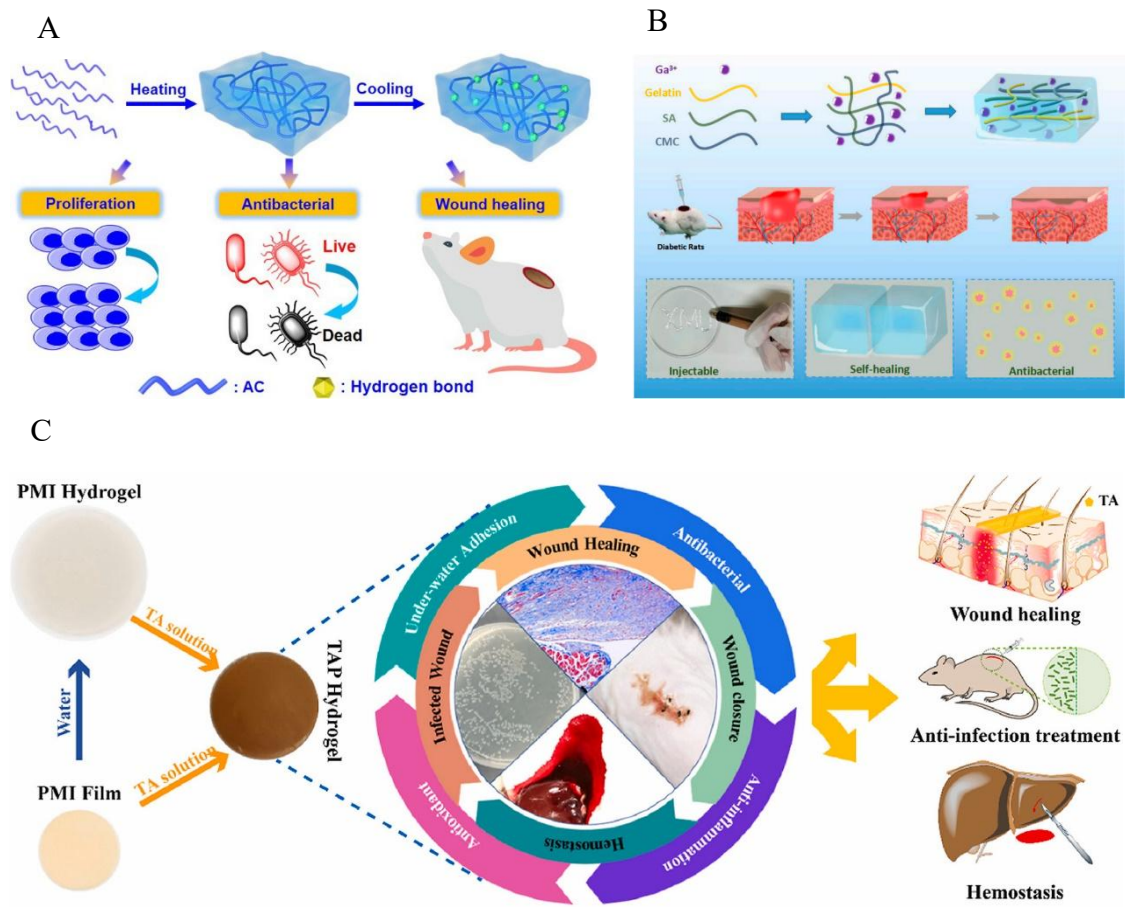

Fig.2 (A) Schematic diagram of AC hydrogel preparation and promotion of wound healing. Copyright@ [61] (B) Preparation, application, and functions of Ga@Gel-Alg-CMC hydrogels. Copyright@ [62] (C) Schematic illustration of the anti-infection, wound healing, incision closure, and hemostasis functions of TAP wound dressing hydrogels. Copyright@[63]

4.1. Role of Hydrogel in Promoting Healing

Hydrogel, as a novel therapeutic material, proves remarkably effective in promoting diabetic wound healing. Yang et al. investigated the use of human umbilical cord

MSC-derived exosomes (hUCMSC-expos) and Pluronic F-127 (PF-127) hydrogel to improve wound healing, showing that the efficient delivery of hUCMSC-expos in PF-127 gel could promote diabetic wound healing [64]. Additionally, xu et al. prepared recombinant human collagen III protein hydrogels with sustained release of extracellular vesicles for skin wound healing[65]. These studies collectively suggest that hydrogels, such as organic-inorganic hybrid hydrogels, SAP hydrogels, and collagen-based hydrogels, have shown promise in promoting diabetic wound healing[61]. Further research is needed to explore the full potential of hydrogels in diabetic wound healing.

The role of hydrogel in promoting diabetic wound healing is mainly manifested in the following aspects: Firstly, the hydrogel can form a transparent gel layer covering the wound surface, maintaining a moist environment, preventing wound dryness and skin cracking, which is conducive to the development and repair of new tissue. Secondly, the active ingredients in hydrogel promote the proliferation and migration of epithelial cells, accelerating the repair and regeneration of the wound epidermis, and advancing the wound healing process. Thirdly, some hydrogels possess anti-inflammatory and antioxidant effects, effectively reducing inflammation around the wound, and decreasing swelling and redness, which is beneficial for wound healing. Finally, hydrogel accelerates the wound healing process, reduces tissue damage, and lowers scar formation, helping to maintain the appearance and function of the wound. In summary, the hydrogel has significant advantages in promoting healing by providing a moist environment, promoting cell growth and migration, suppressing inflammatory reactions, reducing scar formation, accelerating the recovery of diabetic wounds, and improving the quality of life for patients.

4.2. The role of hydrogels in reducing the risk of infection

Diabetic patients are at risk for developing foot ulcers, which can ultimately lead to amputation. Infection is a key contributor to the risk of amputation, and diabetic wounds exhibit an extremely high risk of bacterial infection. The presence of neuropathy, peripheral vascular diseases, and ischemia in diabetic patients leads to the formation of foot ulcerations with a higher risk of infection because the normal response to bacterial infection is missing. Therefore, safe and effective alternative treatments are required to improve diabetic patients' quality of life [66–68]. Hydrogel dressings have been identified as a potential treatment for chronic wound healing in diabetes. Non-healing in wounds for diabetic patients results from a combination of factors that impair the clearing of injured tissue, proliferation of healthy cell populations, and increase the risk of infection. Hydrogel dressings can absorb exudate, prevent infections, and promote wound healing, making them a promising option for the treatment of diabetic foot ulcers [62,69,70]. In addition to their ability to promote wound healing, hydrogels can also serve as a vehicle for delivering therapeutic agents. For example, a novel bone marrow mesenchymal stem cell-derived exosome (MSC-Exo)-loaded hydrogel has been designed for chronic diabetic wound healing, and gallium-doped hydrogels have been shown to slow down wound infection and promote diabetic wound healing [62,71]. Furthermore, pH-modulating hydrogel-based systems have shown promise as a therapeutic strategy for diabetic wounds. Data from preclinical and clinical studies highlight the role of pH in the pathophysiology of diabetic foot ulcers, and topical administration of pH-lowering agents has shown promise for wound healing and infection control in diabetic wounds[72]. In conclusion, hydrogel dressings have the potential to reduce the risk of infection in diabetic wound treatment by promoting wound healing, delivering therapeutic agents, and modulating pH.

4.3. The role of hydrogels in reducing pain and improving quality of life

Diabetic wounds are a significant concern for patients, as they can have a major impact on their quality of life. Factors such as social and economic status, grade of diabetic ulcer, pain, and odor have been found to affect the quality of life of diabetic ulcer patients[73]. In addition to the general wound characteristics, new generations of wound dressings, such as those lasting longer on the wound, can have specific properties such as transferring allogeneic cells to enhance the healing effect and speed up the healing process[74]. Diabetic patients are at risk for developing foot ulcers, which ultimately lead to amputation; hence a safe and effective alternative treatment is required to improve diabetic patients' quality of life [66]. Psychological stress has also been found to have consequences in wound healing, and stress-reducing interventions have been proposed to help patients manage stress and pain while reducing wound inflammation and improving wound healing[75]. Effective strategies are required for the regulation of excessive inflammation and inhibition of MMP-9 overexpression to enhance diabetic wound healing. A bioenergetically-active poly (glycerol sebacate)-based multiblock hydrogel has also been found to improve diabetic wound healing through revitalizing mitochondrial metabolism [76]. In conclusion, diabetic wounds can have a significant impact on a patient's quality of life, and it is important to explore alternative treatments and interventions to improve wound healing and overall well-being. Hydrogels, in particular, have shown promise in improving diabetic wound healing and may offer new opportunities for enhancing the quality of life of diabetic patients.

4.4. Comparison of Hydrogel with Other Treatment Methods:

Traditional Treatment Methods:

Traditional methods for treating diabetic wounds include gauze dressing, topical antibiotics, and surgical debridement. While these methods can control wound infection and promote healing to some extent, they also have limitations. For instance, gauze dressing can lead to wound dryness and bacterial infection, topical antibiotics may cause bacterial resistance, and surgical debridement involves high trauma and cost. In contrast, hydrogel offers advantages such as moisture retention, healing promotion, and antibacterial properties. It's also easy to use, non-invasive, and cost-effective, making it a more ideal treatment choice.

Biological Material Therapy:

Biological material therapy, including biofilms and biological collagen proteins, has emerged as a novel treatment method for wound healing in recent years. These materials exhibit good biocompatibility and tissue absorbability, providing a favorable wound environment to promote healing. However, compared to hydrogel, the effectiveness of biological material therapy may be limited by material source and preparation processes. Additionally, it requires complex use and specialized personnel for operation. In contrast, hydrogel offers a simpler, faster operation process, with a wider range of applications, making it more readily accepted and promoted by clinicians.

Cell Therapy:

Cell therapy, such as stem cell therapy and fibroblast transplantation, is an advanced treatment method that involves transplanting the patient's stem cells or fibroblasts to the wound site to promote tissue regeneration and healing. However, cell therapy faces challenges such as complexity, high cost, and uncertainty in safety and efficacy. In contrast, hydrogel, as a cell-free treatment method, offers a simpler, faster operation process with lower costs, effectively reducing the economic burden on patients and providing a more practical treatment choice.

Other Treatment Methods:

Apart from the above methods, there are other treatment methods such as photodynamic therapy and negative pressure wound therapy. While these methods can promote wound healing to some extent, they also have limitations. For example, photodynamic therapy requires special equipment and professional operation techniques, and negative pressure wound therapy is complex to use and involves high costs. In comparison, hydrogel, as a simple, fast, and cost-effective treatment method, has broader applicability and better accessibility, providing an effective treatment choice for more diabetic patients.

In conclusion, hydrogel offers many advantages in the treatment of diabetic wounds. Compared to traditional treatment methods, it is characterized by simplicity, speed, and low cost, making it likely to become the mainstream treatment method for diabetic wounds. However, further clinical research is needed to validate its safety and efficacy and continuously improve its technology and methods in clinical applications to better serve the treatment needs of diabetic patients.

Hydrogel therapy has shown significant advantages in multiple clinical studies for treating diabetic wounds. Nilforoushzadeh et al. conducted a clinical feasibility study of a novel pre-vascularized skin graft containing the dermal and epidermal layer using the adipose stromal vascular fraction (SVF)--derived endothelial cell population for vascular network regeneration [77]. Yang et al. developed a bioactive skin-mimicking hydrogel band-aid and explored its potential in various medical applications, including diabetic wound healing [63]. Long et al. proposed that MSC-EVs can be combined with 7A to greatly promote diabetic wound healing, potentially constituting a novel therapy for the efficient healing of chronic diabetic wounds [78]. These studies collectively demonstrate the potential of hydrogel therapy in the treatment of diabetic wounds.

Hydrogel therapy has demonstrated significant advantages in multiple clinical studies for treating diabetic wounds. Firstly, it significantly enhances the healing rate of diabetic foot ulcers, expediting the wound healing process. Secondly, hydrogel possesses antimicrobial properties, effectively reducing the risk of wound infection and thereby minimizing the occurrence of complications. Additionally, it alleviates pain sensation, improves patients' quality of life, and enhances sleep quality. Lastly, hydrogel reduces scar formation during wound healing, preserving both the aesthetics and functionality of the wound. These advantages establish hydrogel as a crucial choice for treating diabetic foot ulcers, delivering better treatment outcomes, and improving patients' quality of life.

The application scope of hydrogel in the treatment of diabetic wounds is extensive, covering various types of foot ulcers, including superficial ulcers, deep ulcers, and necrotic tissue. Its practical application is highly convenient, facilitating easy operation in non-hospital settings such as homes and outpatient clinics, thus enhancing treatment accessibility and convenience. However, it's important to note that in specific circumstances, hydrogel may trigger allergic reactions or other adverse responses, necessitating close monitoring of the patient's treatment outcomes and adverse reactions. Overall, as a novel treatment modality, hydrogel holds significant prospects in the treatment of diabetic wounds. With continued research and clinical application of hydrogel technology, its role in the treatment of diabetic wounds is expected to become increasingly prominent, delivering better treatment outcomes and improving patients' quality of life.

5. Discussion

In this review, we have comprehensively discussed and analyzed the application and prospects of hydrogel in the treatment of diabetic wounds. By elaborately discussing hydrogel's role in promoting healing, reducing infection risks, alleviating pain, and improving quality of life, we have reached the following conclusions:

Comprehensive analysis reveals that hydrogel offers significant advantages in the treatment of diabetic wounds: firstly, it effectively promotes wound healing, accelerating tissue repair processes, thus shortening the healing time. Secondly, hydrogel exhibits certain antimicrobial effects, effectively reducing the risk of infection and minimizing the occurrence of complications. Additionally, hydrogel alleviates pain sensation, improves patients' quality of life, and enhances sleep quality. Most importantly, hydrogel therapy significantly improves patients' quality of life, alleviates psychological burdens, and enhances treatment confidence. Therefore, hydrogel demonstrates promising clinical application prospects in the treatment of diabetic wounds, providing a novel and effective means for treating diabetic foot ulcers.

Looking ahead to the future development of hydrogel technology, with the continuous advancement of science and technology, hydrogel technology will undergo continuous improvement to enhance its effectiveness and safety in the treatment of diabetic wounds. As a novel treatment modality, hydrogel is expected to be more widely promoted and applied in clinical practice, providing efficient and safe treatment options for more diabetic patients. Future research should emphasize interdisciplinary collaboration, delving deeper into the mechanisms of hydrogel in the treatment of diabetic wounds, to provide more reliable scientific evidence for its clinical application. In summary, hydrogel, as an emerging treatment modality, holds broad prospects in the treatment of diabetic wounds. We are confident in the future development of hydrogel technology and look forward to its ability to deliver better treatment outcomes and improve the quality of life for diabetic patients.

Declaration of competing interest

The authors declare that they have no competing interests. The authors claim that none of the material in the paper has been published or is under consideration for publication elsewhere. All authors have seen the manuscript and approved to submit to your journal.

*Acknowledgments*

*Fundings:*

Shenzhen Science and Technology Program JCYJ20230807113017035